\documentclass[a4paper]{article}

\usepackage{graphicx}
\usepackage{amsmath}
\usepackage{amsfonts}
\usepackage{hyperref}
\usepackage{xcolor}
\usepackage{cite}
\usepackage{authblk}
\usepackage{epstopdf}

\hypersetup{
     colorlinks   = true,
     citecolor    = blue,
     urlcolor = blue}
\def\Tr{{\rm Tr}}

\def\e{{\rm e}}

\newcommand{\be}{\begin{equation}}
\newcommand{\ee}{\end{equation}}

\newcommand{\bea}{\begin{eqnarray}}
\newcommand{\eea}{\end{eqnarray}}

\newcommand{\ba}{\begin{array}}
\newcommand{\ea}{\end{array}}

\newcommand{\bn}{\begin{enumerate}}
\newcommand{\en}{\end{enumerate}}

\def\Tr{{\rm Tr}}

\def\e{{\rm e}}
\def\GHZ{{\rm GHZ}}
\def\W{{\rm W}}

\newcommand{\ket}[1]{|#1\rangle}

\bibliography{plain}
\title{Entangled graphs: A classification of four-qubit  entanglement}

\author[1]{Masoud Gharahi Ghahi \thanks{E-mail: masoud.gharahi@gmail.com}}
\author[2]{Seyed Javad Akhtarshenas \thanks{E-mail: akhtarshenas@um.ac.ir}}
\affil[1]{\small Department of Physics, Shahid Beheshti University, G.C., Evin, Tehran 19839, Iran}
\affil[2]{\small Department of Physics, Ferdowsi University of Mashhad, Mashhad, Iran}

\begin{document}

\maketitle

\begin{abstract}
We use the concept of \textit{entangled graphs} with
weighted edges to present a classification for four-qubit
entanglement which is based neither on the LOCC nor the SLOCC. Entangled graphs, first introduced by Plesch et al.
[Phys. Rev. A 67, (2003) 012322], are structures such that each
qubit of a multi-qubit system is represented as a vertex and an edge
between two vertices denotes bipartite entanglement between the
corresponding qubits. Our classification is based on the use of
generalized Schmidt decomposition  of pure states of multi-qubit systems. We show that for every possible entangled
graph one can find a pure state such that the reduced entanglement
of each pair, measured by concurrence, represents the weight of the
corresponding edge in the graph. We also use the concept of
tripartite and quadripartite concurrences as a proper measure of
global entanglement of the states. In this case a circle including
the graph indicates the presence of global entanglement.
\end{abstract}



\section{Introduction}
Entanglement, first noticed by Einstein, Podolsky, and Rosen \cite{EPR} and also Schr\"{o}dinger \cite{schr},  is at the heart of quantum mechanics. It turns out that quantum entanglement provides a fundamental potential resource for quantum information tasks such as teleportation and cryptography \cite{barnet}. However, as a resource, entanglement needs to be classified in the sense that each class do the same tasks in quantum information science.

Bipartite pure state entanglement is almost completely understand by
means of Schmidt decomposition \cite{Schmidt}. Schmidt decomposition
allows one to write, by means of local unitary transformation, any
pure state shared by two parties A and B in a canonical form, in the
sense that all the non-local properties of the state are contained
in the positive Schmidt coefficients. Accordingly, any two-qubit
pure state $|\Psi\rangle_2 \in{\mathbb C}^{2}\otimes {\mathbb
C}^{2}$ can be written in the following form \cite{Schmidt} (the
subscript indicates the number of qubits)
\begin{equation}\label{Sch2}
|\Psi\rangle_2=\alpha\ket{00}+\beta\ket{11},\qquad \alpha,\beta\ge
0, \quad \alpha^2+\beta^2=1.
\end{equation}
In this sense any two-qubit pure state is \textit{separable} if
$\alpha\beta=0$, i.e. one of the Schmidt coefficients
$\alpha,\beta$ is zero, or \textit{entangled} if $\alpha\beta\ne
0$.

For three-qubit pure state, there is no a straightforward
generalization of Schmidt decomposition in terms of three orthogonal
product states \cite{Peres1995}. A generalization of two-qubit
Schmidt decomposition for three-qubit pure states has been presented by
Ac\'{i}n et al. \cite{Acin2000}. They showed that for any pure
three-qubit state there exist a local bases which allows one to
build a set of five orthogonal product states in terms of which the
state can be written in a unique form as
\begin{eqnarray}\nonumber\label{Sch3}
\ket{\Psi}_3&=&\lambda_0\ket{000}+\lambda_1\e^{i\phi}\ket{100}+\lambda_2\ket{101}+\lambda_3\ket{110}+\lambda_4\ket{111},
\\ & & \lambda_i\ge 0,\; 0\le \phi\le \pi, \qquad
\sum_{i}\lambda_i^2=1.
\end{eqnarray}
They also presented a canonical form for pure states of
four-qubit system as \cite{Acin2001}
\begin{eqnarray}
|\Psi\rangle_4 \nonumber\label{Sch4}
&=&\alpha|0000\rangle+\beta|0100\rangle+\gamma|0101\rangle+\delta|0110\rangle+\epsilon|1000\rangle+\zeta|1001\rangle
\\ &+&\eta|1010\rangle+\kappa|1011\rangle
+\lambda|1100\rangle+\mu|1101\rangle+\nu|1110\rangle+\omega|1111\rangle.
\end{eqnarray}
The generalized Schmidt decomposition (GSD) for pure states of a general multipartite system is presented by Carteret et al \cite{Carteret2000}.

Multipartite entanglement refers to nonclassical correlations
between three or more quantum particles, and characterization and
quantification of them is far less understood than for  bipartite
entanglement. In the case of multipartite entanglement there are
various aspects of entanglement and, in particular, it is no longer
sufficient to ask \textit{if} the particles are entangled, but one
needs to know \textit{how} they are entangled. There are different
ways that  the multipartite state $\rho$ can be entangled from which
the bipartite entanglement, i.e. the entanglement of the reduced
density matrix $\rho_{kl}$, is only a certain aspect of the
multipartite entanglement properties of $\rho$. For example for
three-qubit states, apart from separable and biseparable pure
states, there exist also two different types of locally inequivalent
genuine tripartite entangled states: the so-called
Greenberger-Horn-Zeilinger (GHZ) type \cite{GHZ1989} and W type
\cite{DVC, AcinBrus2001}, with representatives
$\ket{\GHZ}_3=\frac{1}{\sqrt{2}}(\ket{000}+\ket{111})$ and
$\ket{\W}_3=\frac{1}{\sqrt{3}}(\ket{001}+\ket{010}+\ket{100})$,
respectively. It is well known that states belonging to GHZ and W types cannot be transformed into each other by local operations and classical communication (LOCC). Morover, for three-qubit entanglement, complete classifications were derived \cite{DVC, AcinBrus2001, Sabin2008, kraus}. In \cite{DVC} it has been shown that for three-qubit entanglement, there are six equivalent classes under stochastic local operations and classical communication (SLOCC). However, for four or more qubits, there are infinite SLOCC classes  \cite{lili}, and it is therefore highly desirable to partition the infinite
classes into a finite number of families. In \cite{VDDV}, Verstraete et al. have shown that there are nine families in four-qubit entanglement. In addition, many attempts have been taken for classification of four-qubit entanglement in complete sets \cite{lamata, cao, li, borsten} or in special cases \cite{park1, park2}. So, it seems that four-qubit entanglement still requires further study.

The entanglement properties of a multi-qubit system may be
represented mathematically in several ways. In \cite{Dur2001},
D\"{u}r has investigated the entanglement properties of multipartite
systems, concentrating on the bipartite aspects of multipartite
entanglement, i.e. the bipartite entanglement that is robust against
the disposal of the particles. In this sense, the two qubits
belonging to a multipartite system are entangled if their reduced
density matrix is entangled. D\"{u}r introduces  the concept of
\textit{entanglement molecules}, i.e. structures such that each
qubit is represented by  an atom while an entanglement between a
pair of qubits is represented by a bound. He also shows that an
arbitrary entanglement molecule can be represented by a mixed state
of a multiqubit system. On the other hand, Plesch et al.
\cite{Buzek2003,Buzek2004} introduce the concept of
\textit{entangled graphs} such that each qubit of a multipartite
system is represented as a vertex and an edge between two vertices
denotes bipartite entanglement between the corresponding qubits.
They show that any entangled graph of $N$ qubits can be
represented by a pure state from a subspace of the whole
$2^N$-dimensional Hilbert space of N qubits.

In this paper we turn our attention to entangled graphs and
associate a pure four-qubit state to each graph of order 4. This,
naturally,  leads to a classification of entanglement in four-qubit
systems. Our classification is based on the use of GSD of pure states of multi-qubit systems. We show that for every possible entangled graph one can find a pure state such that the reduced
entanglement of each pair, measured by concurrence \cite{Wootters}, represents the
weight of the corresponding edge in the graph.  In order to characterize the global entanglement of the state, we use the
concept of tripartite and quadripartite concurrences, introduced in
\cite{Love2007}, in such a way that  a circle including the graph
indicates the nonzero global entanglement of the corresponding
state. It is worth to mention that our classification does not contain permutations.

The paper is organized as follows: In section 2 we present the
tripartite concurrence and give a classification of the tripartite
pure states, Section 3 is devoted to present a classification for
four-qubit pure states. We conclude the paper in section 4, with a
brief conclusion.

\section{Classification of three-qubit entanglement}
\begin{figure}[t]
\centerline{\includegraphics[height=8cm]{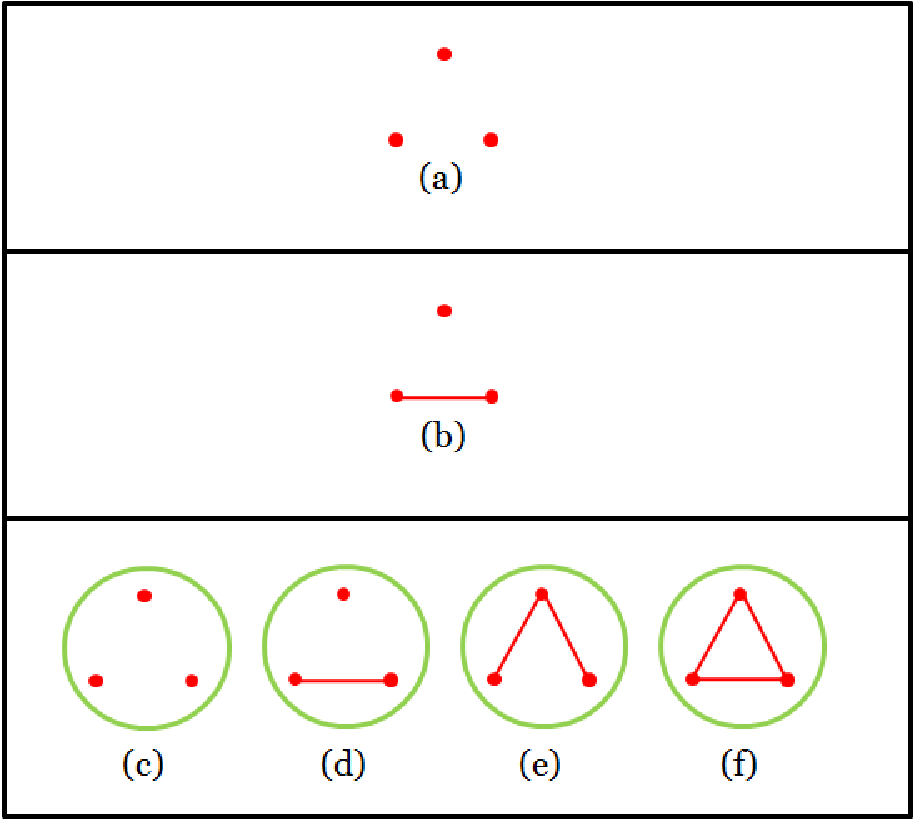}} \caption{Three-qubit
entangled graphs and classification of entanglement in three-qubit
system. (a) Fully separable, (b) Biseparable, (c)-(f) Fully
inseparable.}
\label{figure:1}
\end{figure}

In this section we provide a classification for three-qubit pure states, and associate to each possible entangled state an entangled graph (see Figure \ref{figure:1}).  Entangled graphs are structures such
that each qubit  is represented as a vertex and an edge between two
vertices denotes bipartite entanglement between the corresponding
qubits. Accordingly, an edge connecting two vertices indicates that the entanglement between the corresponding qubits is robust against the disposal of the remaining qubits. On the other hand, following  \cite{Sabin2008},  we use a circle
including three vertices in order to indicate full tripartite
entanglement. We show that for every possible entangled graph one
can find a pure state such that the reduced entanglement of each
pair, measured by concurrence, represents the weight of the
corresponding edge in the graph. In order to characterize the full
entanglement of the state we use the tripartite concurrence defined
by \cite{Love2007}
\begin{equation}\label{C123}
{\mathcal{C}}_{123}(\rho)=\left({\mathcal{C}}_{1(23)}{\mathcal{C}}_{2(13)}{\mathcal{C}}_{3(12)}
\right)^{\frac{1}{3}},
\end{equation}
where the bipartite concurrence ${\mathcal{C}}_{A(BC)}$ is defined as
${\mathcal{C}}_{A(BC)}=\sqrt{2(1-\Tr(\rho^2_{A}))}$ where $\rho_A$ is the
reduced density matrix obtained from the whole density matrix $\rho$
by taking trace over the two subsystems B and C. For a general two-qubit pure state $|\psi\rangle=a|00\rangle+b|01\rangle+c|10\rangle+d|11\rangle$, the concurrence takes the simple form ${\mathcal{C}}=2|ad-bc|$.

\begin{table}[ht] 
\caption{Three-qubit GSD's coefficients that make nonzero bipartite
concurrence} 
\centering
\begin{tabular}{|c|c|c|}
\hline

& & \\ [-1.5ex]

${\mathcal{C}}_{12}$ & ${\mathcal{C}}_{13}$ & ${\mathcal{C}}_{23}$ \\ [1ex]

\hline

& & \\ [-1.5ex]

($\lambda_{0}$ , $\lambda_{3}$) & ($\lambda_{0}$ , $\lambda_{2}$) &
($\lambda_{1}$ , $\lambda_{4}$) \\ [1ex]

& & ($\lambda_{2}$ , $\lambda_{3}$)\\ [1ex]

\hline

\end{tabular}
\label{table:1}
\end{table}

In order to present our classification for three-qubit states, let us first calculate the bipartite concurrence of the GSD of three-qubit (\ref{Sch3}). We find ${\mathcal{C}}_{12}=2\lambda_{0} \lambda_{3}$,
${\mathcal{C}}_{13}=2\lambda_{0} \lambda_{2}$ and ${\mathcal{C}}_{23}=2\sqrt{(\lambda_{1}\lambda_{4})^{2}+(\lambda_{2}\lambda_{3})^{2}-2(\lambda_{1}\lambda_{4})^{2}(\lambda_{2}\lambda_{3})^{2}\cos{\phi}}$, which reduces to ${\mathcal{C}}_{23}=2|\lambda_{1}\lambda_{4}-\lambda_{2}\lambda_{3}|$ if we choose $\phi=0$.
Now, let us consider all possible bipartite factorizations with nonzero concurrence of the three-qubit GSD state (\ref{Sch3}). This enables us to find  coefficients of the GSD which make nonzero bipartite concurrences and guarantee a weighted edge between arbitrary vertices. Moreover, in order to have nonzero global entanglement we must choose proper Schmidt coefficients such that none of the qubits can be factorized  from the state. These considerations are summarized in  Table \ref{table:1}. It follows from this table that in order to have nonzero two-qubit concurrence  ${\mathcal{C}}_{12}$, we have to choose a nonzero value for the pair  $(\lambda_0, \lambda_3)$, but for nonzero ${\mathcal{C}}_{23}$ we are allowed to choose either $(\lambda_1, \lambda_4)$ or $(\lambda_2, \lambda_3)$ to be nonzero. Now, our classification is as follows:

\begin{enumerate}
\item \textbf{Fully separable:} A pure state  $\ket{\psi}$ is
fully separable or unentangled if it can be written as
$\ket{A}_{3}=\ket{\phi_{1}}\otimes\ket{\phi_{2}}\otimes\ket{\phi_{3}}$.
The corresponding graph has three vertices without any edge. All
bipartite entanglements as well as the global entanglement of this
state are zero, i.e. ${\mathcal{C}}_{12}={\mathcal{C}}_{13}={\mathcal{C}}_{23}=0$ and ${\mathcal{C}}_{123}=0$. These states can be represented by a graph of order 3, without any edge
(see Fig. \ref{figure:1}(a)).  Any state which cannot be written in fully
separable form is entangled and belongs to one of the following
categories.

\item \textbf{Biseparable:} A pure state  $\ket{\psi}$ is
biseparable if it is not fully separable and that it can be written
as $\ket{\phi_{ij}}\otimes\ket{\phi_{k}}$ for one $k\in\{1,2,3\}$.
A representative of this class is given by
\begin{equation}
|B\rangle_{3}=\left(\lambda_{0}|00\rangle+\lambda_{3}|11\rangle\right)|0\rangle.
\end{equation}
In this case we find for bipartite concurrences ${\mathcal{C}}_{12}=2\lambda_{0}
\lambda_{3}$, ${\mathcal{C}}_{13}={\mathcal{C}}_{23}=0$. This state does not have global
entanglement and therefore we have ${\mathcal{C}}_{123}=0$. The corresponding
graph has three vertices with only one edge connecting two vertices
$i$ and $j$ (see Fig. \ref{figure:1}(b)).

\item \textbf{Fully inseparable:} A pure state $\ket{\psi}$ is
fully inseparable if it is not separable or biseparable. A fully
inseparable state has nonzero global entanglement, i.e.
${\mathcal{C}}_{123}\ne 0$. This category includes the following classes (see
Figs. \ref{figure:1}(c)-(f)):

\begin{enumerate}
\item \textit{Three vertices without any edge but a circle includes the graph (Fig. \ref{figure:1}(c)):} A pure representative of this class is as follows
\begin{equation}
|C\rangle_{3}=\lambda_{0}|000\rangle+\lambda_{4}|111\rangle.
\end{equation}
Surprisingly, similar to fully separable states, all bipartite
entanglements of this state are zero, i.e. ${\mathcal{C}}_{12}={\mathcal{C}}_{13}={\mathcal{C}}_{23}=0$,
but the state has nonzero global entanglement, i.e.
${\mathcal{C}}_{123}=2\lambda_{0}\lambda_{4}$. This class corresponds to GHZ states.

\item \textit{Three vertices with one edge and a circle includes the graph (Fig. \ref{figure:1}(d)):} This class has the following representative
\begin{equation}
|D\rangle_{3}=\lambda_{0}|000\rangle+\lambda_{3}|110\rangle+\lambda_{4}|111\rangle.
\end{equation}
In this case for bipartite concurrences we find ${\mathcal{C}}_{12}=2\lambda_{0}
\lambda_{3}$, ${\mathcal{C}}_{13}={\mathcal{C}}_{23}=0$ and for global entanglement we get
${\mathcal{C}}_{123}=2\lambda_{0}\left(\lambda_{4}(\lambda_{3}^2+\lambda_{4}^2)\right)^\frac{1}{3}$.

\item \textit{Three vertices with  two edges and a circle includes the graph (Fig. \ref{figure:1}(e)):} This class is represented by the
following state
\begin{equation}
|E\rangle_{3}=\lambda_{0}|000\rangle+\lambda_{1}|100\rangle+\lambda_{3}|110\rangle+\lambda_{4}|111\rangle.
\end{equation}
The state has two nonzero bipartite concurrences as
${\mathcal{C}}_{12}=2\lambda_{0} \lambda_{3}$ and ${\mathcal{C}}_{23}=2\lambda_{1}
\lambda_{4}$. The nonzero global entanglement of the state is given by \\ ${\mathcal{C}}_{123}=2\left(\sqrt{\lambda_{0}^2(\lambda_{3}^2+\lambda_{4}^2)}\sqrt{\lambda_{0}^2(\lambda_{3}^2+\lambda_{4}^2)
+\lambda_{1}^2\lambda_{4}^2}\sqrt{\lambda_{4}^2(\lambda_{0}^2+\lambda_{1}^2)}\right)^\frac{1}{3}
$.

\item \textit{Three vertices with  three edges and a circle includes the graph (Fig. \ref{figure:1}(f)):} The last class of this category has the
following representative
\begin{equation}
|F\rangle_{3}=\lambda_{0}|000\rangle+\lambda_{2}|101\rangle+\lambda_{3}|110\rangle.
\end{equation}
All bipartite concurrences of this state are nonzero as
${\mathcal{C}}_{12}=2\lambda_{0} \lambda_{3}$, ${\mathcal{C}}_{13}=2\lambda_{0} \lambda_2{}$
and ${\mathcal{C}}_{23}=2\lambda_{2} \lambda_{3}$. The state has also a nonzero global entanglement as ${\mathcal{C}}_{123}=2\left(\sqrt{\lambda_{0}^2(\lambda_{2}^2+\lambda_{3}^2)}\sqrt{\lambda_{3}^2(\lambda_{0}^2+\lambda_{2}^2)}
\sqrt{\lambda_{2}^2(\lambda_{0}^2+\lambda_{3}^2)}\right)^\frac{1}{3}$.
This class corresponds to W states.

In \cite{kraus}, the authors derived a decomposition for three-qubit pure states, which, as the Schmidt decomposition for bipartite states, can be easily computed. Their work led to a classification beyond the SLOCC paradigm, since there exist three classes correspond to GHZ-type states. If we use tangle $\tau_{123}$ \cite{CKW}, we find that classes in Fig. \ref{figure:1}(c)-(e) are GHZ-type and the class in Fig. \ref{figure:1}(f) is W-type.

\end{enumerate}
\end{enumerate}

\section{Classification of four-qubit entanglement}
This section is devoted to provide a classification for four-qubit pure states and associate to each state an entangled graph (see Figure \ref{figure:2}). In order to characterize the global
entanglement of the state we use the quadripartite concurrence defined
by \cite{Love2007}
\begin{equation}\label{C1234}
{\mathcal{C}}_{1234}(\rho)=\left({\mathcal{C}}_{1(234)}{\mathcal{C}}_{2(134)}{\mathcal{C}}_{3(124)}{\mathcal{C}}_{4(123)}{\mathcal{C}}_{(12)(34)}{\mathcal{C}}_{(13)(24)}{\mathcal{C}}_{(14)(23)}\right)^{\frac{1}{7}},
\end{equation}
where ${\mathcal{C}}_{A(BCD)}$ and ${\mathcal{C}}_{AB(CD)}$ are defined as
${\mathcal{C}}_{A(BCD)}=\sqrt{2(1-\Tr(\rho^2_{A}))}$ and
${\mathcal{C}}_{(AB)(CD)}=\sqrt{\frac{4}{3}(1-\Tr(\rho^2_{AB}))}$, respectively.

In a similar manner as in three-qubit case, let us consider all possible bipartite factorizations with nonzero  pairwise concurrences of the four-qubit GSD state (\ref{Sch4}). Table \ref{table:2} summarizes this in such a way that every column of the Table corresponds to a pair of vertices $i$ and $j$ and represents all possible pairs of coefficients which make nonzero concurrence between the vertices $i$ and $j$.    For a nonzero global entanglement, one can choose proper coefficients that no qubit could be factorized. This is somewhat difficult but feasible. Using these considerations, our four-qubit classification is as follows:

\begin{figure}[t]
\centerline{\includegraphics[height=12.5cm]{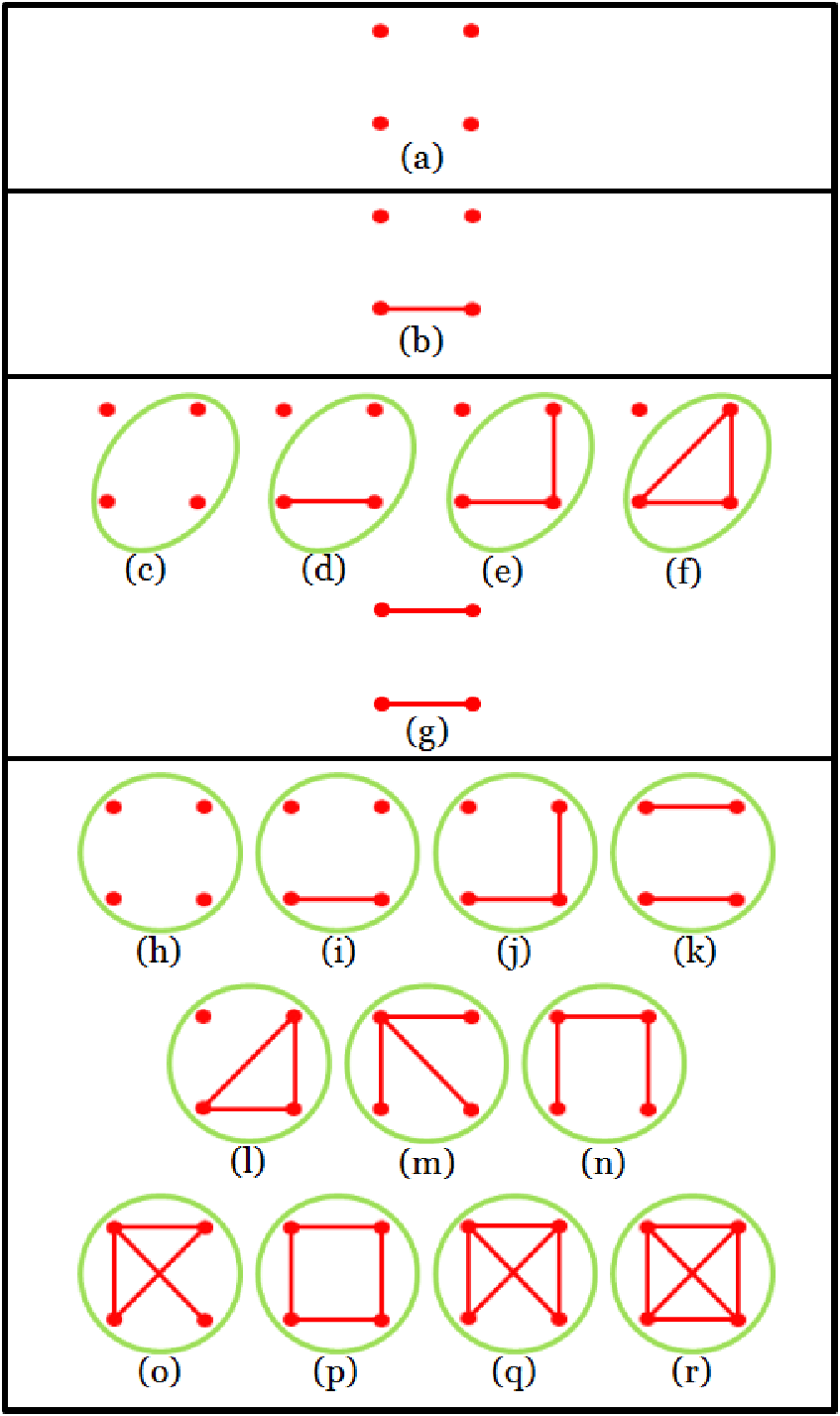}} \caption{Four-qubit
entangled  graphs and classification of entanglement in four-qubit
system. (a) Fully separable, (b) Tri-separable, (c)-(g) Biseparable,
(h)-(r) Fully inseparable.}
\label{figure:2}
\end{figure}

\begin{table}[ht]
\caption{Four-qubit GSD's coefficients that make nonzero bipartite
concurrence}
\centering
\begin{tabular}{|c|c|c|c|c|c|}
\hline

& & & & & \\ [-1.5ex]

${\mathcal{C}}_{12}$ & ${\mathcal{C}}_{13}$ & ${\mathcal{C}}_{14}$ & ${\mathcal{C}}_{23}$ & ${\mathcal{C}}_{24}$ & ${\mathcal{C}}_{34}$ \\
[1ex]

\hline

& & & & & \\ [-1.5ex]

($\alpha$ , $\lambda$) & ($\alpha$ , $\eta$) & ($\alpha$ , $\zeta$) &
($\alpha$ , $\delta$) & ($\alpha$ , $\gamma$) & ($\epsilon$ , $\kappa$) \\ [1ex]

($\beta$ , $\epsilon$) & ($\beta$ , $\nu$) & ($\beta$ , $\mu$) &
($\epsilon$ , $\nu$) & ($\epsilon$ , $\mu$) & ($\lambda$ , $\omega$) \\ [1ex]

($\gamma$ , $\zeta$) & ($\gamma$ , $\omega$) & ($\delta$ , $\omega$)
& ($\zeta$ , $\omega$) & ($\eta$ , $\omega$) & ($\gamma$ , $\delta$) \\ [1ex]

($\delta$ , $\eta$) & ($\delta$ , $\lambda$) & ($\gamma$ , $\lambda$)
& ($\eta$ , $\lambda$) & ($\zeta$ , $\lambda$) & ($\zeta$ , $\eta$) \\ [1ex]

& & & ($\kappa$ , $\mu$) & ($\kappa$ , $\nu$) & ($\mu$ , $\nu$) \\ [1ex]

\hline

\end{tabular}
\label{table:2}
\end{table}

\begin{enumerate}
\item \textbf{Fully separable:} A pure state  $\ket{\psi}$ is
fully separable if it can be written as
$\ket{\phi_{1}}\otimes\ket{\phi_{2}}\otimes\ket{\phi_{3}}\otimes\ket{\phi_{4}}$.
All bipartite entanglements as well as the tripartite and global
entanglement of this state are zero, i.e. ${\mathcal{C}}_{ij}=0$, ${\mathcal{C}}_{ijk}=0$
and ${\mathcal{C}}_{1234}=0$. The corresponding graph has four vertices without
any edge (see Fig. \ref{figure:2}(a)).  Any state which cannot be written in
fully separable form, is entangled and belongs to one of the
following categories.

\item \textbf{Tri-separable:} A pure state  $\ket{\psi}$ is
tri-separable if it is not fully separable but it can be written as
$\ket{\phi_{ij}}\otimes\ket{\phi_{k}}\otimes\ket{\phi_{l}}$ for
$k,l\in\{1,2,3,4\}$. A representative for this class is given by
\begin{equation}
|B\rangle_{4}=\left(\alpha|00\rangle+\lambda|11\rangle\right)\ket{00}.
\end{equation}
Only one bipartite concurrence of this state is nonzero, i.e.
${\mathcal{C}}_{12}=2\alpha \lambda$ and ${\mathcal{C}}_{13}={\mathcal{C}}_{14}={\mathcal{C}}_{23}={\mathcal{C}}_{24}={\mathcal{C}}_{34}=0$.
This state also has a zero tripartite entanglement ${\mathcal{C}}_{ijk}=0$ and a
zero global entanglement ${\mathcal{C}}_{1234}=0$. The corresponding graph of
this category has four vertices with only one edge (see Fig. \ref{figure:2}(b)).

\item \textbf{Biseparable} A pure state  $\ket{\psi}$ is
bi-separable if it is not tri-separable and that it can be written
as $\ket{\phi_{ijk}}\otimes\ket{\phi_{l}}$ for one $l\in\{1,2,3,4\}$
or as $\ket{\phi_{ij}}\otimes\ket{\phi_{kl}}$ for one
$kl\in\{12,13,14,23,24,34\}$. For the former, the global
entanglement is zero and the following classes are defined (see Figs
\ref{figure:2}(c)-(g)):
\begin{enumerate}
\item \textit{Four vertices without any edge, but a circle includes three of vertices (Fig. \ref{figure:2}(c)):}
 A pure state representative of this class is as
\begin{equation}
|C\rangle_{4}=|1\rangle\left(\epsilon|000\rangle+\omega|111\rangle\right).
\end{equation}
All bipartite concurrences as well as global entanglement of this
state are zero, i.e.  ${\mathcal{C}}_{ij}=0$ and ${\mathcal{C}}_{1234}=0$, but the system
has  entanglement through three vertices 2,3,4, i.e. ${\mathcal{C}}_{234}\neq0$.

\item \textit{Four vertices with only one edge and a circle includes three of vertices (Fig. \ref{figure:2}(d)):}
 The pure state representative of this class is as
\begin{equation}
|D\rangle_{4}=|1\rangle\left(\epsilon|000\rangle+\nu|110\rangle+\omega|111\rangle\right).
\end{equation}
This state has zero value for all but one  bipartite concurrences as
${\mathcal{C}}_{23}=2\epsilon \nu$. Also the global entanglement of the state is
zero but the entanglement through three vertices 2,3,4, is nonzero,
i.e. ${\mathcal{C}}_{1234}=0$, ${\mathcal{C}}_{234}\neq0$.

\item \textit{Four vertices with two  edges and a circle includes three of vertices (Fig. \ref{figure:2}(e)):}
 The pure state representative of this class is as
\begin{equation}
|E\rangle_{4}=|1\rangle\left(\epsilon|000\rangle+\lambda|100\rangle+\nu|110\rangle+\omega|111\rangle\right).
\end{equation}
The nonzero bipartite concurrences of this state are
${\mathcal{C}}_{23}=2\epsilon \nu$, ${\mathcal{C}}_{34}=2\lambda \omega$. The state also has zero global entanglement, i.e. ${\mathcal{C}}_{1234}=0$, but nonzero tripartite entanglement ${\mathcal{C}}_{234}\neq0$.

\item \textit{Four vertices with three edges and a circle includes three of vertices (Fig. \ref{figure:2}(f)) :}  The pure state representative of this
class is as
\begin{equation}
|F\rangle_{4}=|1\rangle\left(\epsilon|000\rangle+\mu|101\rangle+\nu|110\rangle\right).
\end{equation}
The nonzero bipartite concurrences of this state are as
${\mathcal{C}}_{23}=2\epsilon \nu$, ${\mathcal{C}}_{24}=2\epsilon \mu$, ${\mathcal{C}}_{34}=2\mu
\nu$. Again this state has a zero global entanglement ${\mathcal{C}}_{1234}=0$
and a nonzero tripartite entanglement ${\mathcal{C}}_{234}\neq0$.

\item \textit{Four vertices with two edges and no circle (Fig. \ref{figure:2}(g)) :}  The pure state representative of this
class is as
\begin{equation}
|G\rangle_{4}= \gamma|0101\rangle+\delta|0110\rangle+\zeta|1001\rangle+\eta|1010\rangle\equiv(a|01\rangle+b|10\rangle)_{12}\otimes (c|01\rangle+d|10\rangle)_{34}.
\end{equation}
The nonzero bipartite concurrences of this state are as
${\mathcal{C}}_{12}=2(\gamma \zeta+\delta \eta)$ and ${\mathcal{C}}_{34}=2(\gamma \delta+
\zeta \eta)$. This state has a zero global entanglement
${\mathcal{C}}_{1234}=0$.
\end{enumerate}

\item \textbf{Fully inseparable} A pure state $\ket{\psi}$ is
fully inseparable if it is  not separable, tri-separable or
biseparable. This category has a nonzero global entanglement, i.e.
$C_{1234}\neq 0$, and  has the following classes (see Figs.
\ref{figure:2}(h)-(r)):
\begin{enumerate}
\item \textit{Four vertices without any edge, but a circle includes the graph (Fig. \ref{figure:2}(h)):}  A pure state representative of this class is as
\begin{equation}
|H\rangle_{4}=\alpha|0000\rangle+\omega|1111\rangle.
\end{equation}
All bipartite concurrences of this state are zero, i.e. ${\mathcal{C}}_{ij}=0$,
but the state has a nonzero global entanglement
${\mathcal{C}}_{1234}=2\alpha\omega$.
This class corresponds to GHZ states.

\item \textit{Four vertices with one edge and a circle includes the graph (Fig. \ref{figure:2}(i)):}
This class has the following representative
\begin{equation}
|I\rangle_{4}=\alpha|0000\rangle+\kappa|1011\rangle+\mu|1101\rangle.
\end{equation}
The only nonzero bipartite concurrence is ${\mathcal{C}}_{23}=2\kappa \mu$. Also
the state has a nonzero global entanglement ${\mathcal{C}}_{1234}\ne 0$.

\item \textit{Four vertices with two edges and a circle includes the graph (Fig. \ref{figure:2}(j)):}
This class has the following representative
\begin{equation}
|J\rangle_{4}=\alpha|0000\rangle+\eta|1010\rangle+\kappa|1011\rangle+\mu|1101\rangle.
\end{equation}
In this case two nonzero bipartite concurrences are ${\mathcal{C}}_{13}=2\alpha
\eta$ and ${\mathcal{C}}_{23}=2\kappa \mu$. The state also has a nonzero global
entanglement ${\mathcal{C}}_{1234}\neq0$.
\item \textit{Four vertices with two edges and a circle includes the graph (Fig. \ref{figure:2}(k)):}
This class has the following representative
\begin{equation}
|K\rangle_{4}=\alpha|0000\rangle+\eta|1010\rangle+\omega|1111\rangle.
\end{equation}
The state has two nonzero bipartite concurrences as ${\mathcal{C}}_{13}=2\alpha
\eta$ and ${\mathcal{C}}_{24}=2\eta \omega$, and a nonzero
global entanglement ${\mathcal{C}}_{1234}\ne 0$.

\item \textit{Four vertices with three edges and a circle includes the graph (Fig. \ref{figure:2}(l)):}
This class has the following representative
\begin{equation}
|L\rangle_{4}=\gamma|0101\rangle+\kappa|1011\rangle+\mu|1101\rangle+\nu|1110\rangle.
\end{equation}
The state has three nonzero bipartite concurrences as
${\mathcal{C}}_{23}=2\kappa \mu$, ${\mathcal{C}}_{24}=2\kappa \nu$ and ${\mathcal{C}}_{34}=2\mu \nu$.
This state also has a nonzero global entanglement ${\mathcal{C}}_{1234}\ne 0$.

\item \textit{Four vertices with three edges and a circle includes the graph (Fig. \ref{figure:2}(m)):}
This class has the following representative
\begin{equation}
|M\rangle_{4}=\alpha|0000\rangle+\beta|0100\rangle+\lambda|1100\rangle+\mu|1101\rangle+\nu|1110\rangle+\omega|1111\rangle, \quad \mu \nu=\lambda \omega.
\end{equation}
Three nonzero bipartite concurrences are as ${\mathcal{C}}_{12}=2\alpha \lambda$, ${\mathcal{C}}_{13}=2\beta \nu$, and
${\mathcal{C}}_{14}=2\beta \mu$. The state also has a nonzero global
entanglement ${\mathcal{C}}_{1234}\ne 0$. If $\mu\nu\neq\lambda\omega$, so ${\mathcal{C}}_{34}=2|\mu\nu-\lambda\omega|$ and the state describes Fig. \ref{figure:2}(o).

\item \textit{Four vertices with three edges and a circle includes the graph (Fig. \ref{figure:2}(n)):}
This graph has the following pure state representative
\begin{equation}
|N\rangle_{4}=\alpha|0000\rangle+\kappa|1011\rangle+\lambda|1100\rangle+\mu|1101\rangle+\omega|1111\rangle.
\end{equation}
This state has three nonzero bipartite concurrences as ${\mathcal{C}}_{12}=2\alpha\lambda$, ${\mathcal{C}}_{23}=2\kappa \mu$ and ${\mathcal{C}}_{34}=2\lambda \omega$. Again the state has a nonzero global entanglement ${\mathcal{C}}_{1234}\neq0$.

\item \textit{Four vertices with four edges and a circle includes the graph (Fig. \ref{figure:2}(o)):}
This class has the following representative
\begin{equation}
|O\rangle_{4}=\alpha|0000\rangle+\zeta|1001\rangle+\kappa|1011\rangle+\lambda|1100\rangle+\mu|1101\rangle.
\end{equation}
This state has four nonzero bipartite concurrences as ${\mathcal{C}}_{12}=2\alpha \lambda$, ${\mathcal{C}}_{14}=2\alpha \zeta$, ${\mathcal{C}}_{23}=2\kappa
\mu$ and ${\mathcal{C}}_{24}=2\zeta \lambda$.  The state also has a nonzero global entanglement ${\mathcal{C}}_{1234}\neq0$.

\item \textit{Four vertices with four edges and a circle includes the graph (Fig. \ref{figure:2}(p)):}
This class has the following representative
\begin{equation}
|P\rangle_{4}=\alpha|0000\rangle+\zeta|1001\rangle+\eta|1010\rangle+\omega|1111\rangle, \quad \alpha \omega \geq 2\zeta \eta \quad
\textmd{or} \quad \alpha \omega=\zeta \eta.
\end{equation}
Four nonzero bipartite concurrences are as ${\mathcal{C}}_{13}=2\alpha\eta$, ${\mathcal{C}}_{14}=2\alpha\zeta$, ${\mathcal{C}}_{23}=2\zeta\omega$ and ${\mathcal{C}}_{24}=2\eta\omega$. The state also has a nonzero global entanglement ${\mathcal{C}}_{1234}\neq0$. If $\zeta\eta>\alpha\omega$, so ${\mathcal{C}}_{34}=2(\zeta\eta-\alpha\omega)$ and the state describes Fig. \ref{figure:2}(q).

\item \textit{Four vertices with five edges and a circle includes the graph (Fig. \ref{figure:2}(q)):}
This class has the following representative
\begin{equation}
|Q\rangle_{4}=\alpha|0000\rangle+\epsilon|1000\rangle+\eta|1010\rangle+\lambda|1100\rangle+\mu|1101\rangle+\nu|1110\rangle, \quad \epsilon \nu \neq \eta \lambda.
\end{equation}
In this case we find five nonzero bipartite concurrences as ${\mathcal{C}}_{12}=2\alpha\lambda$,
${\mathcal{C}}_{13}=2\alpha\eta$, ${\mathcal{C}}_{23}=2|\epsilon\nu - \eta\lambda|$, ${\mathcal{C}}_{24}=2\epsilon\mu$ and ${\mathcal{C}}_{34}=2\mu\nu$.  The state also has a nonzero global entanglement ${\mathcal{C}}_{1234}\neq0$. It is obvious that if $\epsilon\nu=\eta\lambda$, the state describes Fig. \ref{figure:2}(p).

\item \textit{Four vertices with six edges and a circle includes the graph (Fig. \ref{figure:2}(r)):}
This class has the following representative
\begin{equation}
|R\rangle_{4}=\alpha|0000\rangle+\zeta|1001\rangle+\eta|1010\rangle+\lambda|1100\rangle.
\end{equation}
All bipartite concurrences of this state are nonzero and given by ${\mathcal{C}}_{12}=2\alpha
\lambda$, ${\mathcal{C}}_{13}=2\alpha \eta$, ${\mathcal{C}}_{14}=2\alpha \zeta$,
${\mathcal{C}}_{23}=2\eta \lambda$, ${\mathcal{C}}_{24}=2\zeta \lambda$ and ${\mathcal{C}}_{34}=2\zeta
\eta$.  This state also has a nonzero global entanglement ${\mathcal{C}}_{1234}\ne
0$. This class corresponds to W states.
\end{enumerate}
\end{enumerate}

\section{Conclusions}
We have used the concept of entangled graphs with weighted edges and
have presented a classification of three-qubit and four-qubit
entanglement.  For  pure states,  each qubit of
a multi-qubit system is represented by a vertex and an edge between
two vertices denotes bipartite entanglement between the
corresponding qubits. The nonzero concurrence of the two-qubit
reduced density matrix indicates that the entanglement of the state
is robust against the disposal of remaining particles and have
represented by an edge connecting the corresponding vertices. In
this classification, we have used the generalized Schmidt
decomposition. It is shown that for every possible entangled graph,
one can find a pure state in minimal form, i.e. with minimum number of coefficients, such that the reduced entanglement of each pair, measured by concurrence, represents the weight of the
corresponding edge in the graph. By using the concept of
tripartite and quadripartite concurrences, we have characterized the
global entanglement of the tripartite and quadripartite pure states,
respectively. The presence of global entanglement of the pure state is represented by a circle including the corresponding graph. To sum up, we have classified entanglement according to all possible entanglement structure in terms of entangled graphs without considering permutations. It is obvious that there are many states in each class, considering permutations. Moreover, by computing bipartite and global entanglements, it is possible to decide which class an arbitrary state belongs to. We think this paradigm would be extendable for N-qubit entanglement. In addition, it is worth to mention that our approach is convenient to write a pure state for an arbitrary entangled graph without considering classification, which would be fruitful for certain tasks in quantum information science.

\section*{Acknowledgements}
This work was partially performed at the Department of Physics, University of Isfahan.

\newpage

\end{document}